\documentclass[
    ,final            
  ]
  {aipproc}

\layoutstyle{6x9}

\begin{document}

\title{Double Longitudinal Spin Asymmetry in Neutral Pion Production in Polarized
 p+p Collisions at $\sqrt{s}=200$ GeV at PHENIX}

\classification{14.20.Dh, 13.85.Ni, 13.88.+e}

\keywords{Proton structure, Proton spin}

\author{Kieran Boyle on behalf of the PHENIX collaboration}{
  address={Department of Physics and Astronomy, Stony Brook University, Stony Brook, NY, 11794}
}

\begin{abstract}
A major goal of the RHIC spin program is to measure $\Delta g$, the gluon
contribution to the proton's spin. Measurements by PHENIX of the double
longitudinal spin asymmetry, $A_{LL}$, of the neutral pion production
at mid-rapidity in polarized proton collisions have been shown to
constrain $\Delta g$. Results from the 2005 RHIC run, as well as high
$p_T$ data from the 2006 RHIC run, are presented.  The results disfavor
maximal positive and negative values of $\Delta g$.  A measurement of
azimuthally independent double transverse spin asymmetry, $A_{TT}$, is
also presented.
\end{abstract}

\maketitle


\section{Introduction}

In 1988, using polarized DIS, the European Muon Collaboration found
$\Delta \Sigma$, the quark contribution to the proton spin, to be
consistent with zero. Since then, many experiments have shown that
while nonzero, $\Delta \Sigma$ ($\sim$25-35\%) is smaller than
expected by a naive quark model. The remaining component must be
carried by the gluon spin and the quark and gluon orbital angular
momentum.

The gluon contribution is not well constrained by these fixed target
experiments.  By colliding longitudinally polarized protons at the
Relativistic Heavy Ion Collider (RHIC) at Brookhaven National
Laboratory (BNL), we can probe the gluon contribution to the proton
spin directly.  For any specified interaction, $p+p\rightarrow C+X$,
by measuring the asymmetry in the production of C between like and
unlike helicity collisions, we can probe the gluon contribution to
the proton spin, $\Delta g$.  This double longitudinal spin
asymmetry is defined as
\begin{equation}
A_{LL} = \frac{\sigma_{++} - \sigma_{+-}}{\sigma_{++} + \sigma_{+-}}
\label{A_LL crosssection}
\end{equation}
where $\sigma_{++}$ ($\sigma_{+-}$) is the cross section with same
(opposite) helicity.  Experimentally, what is actually measure is
\begin{equation}
A_{LL} = \frac{1}{|P_1||P_2|} \frac{N_{++} - RN_{+-}}{N_{++} +
RN_{+-}},\quad R=\frac{L_{++}}{L_{+-}}.
\label{A_LL_RL}
\end{equation}
where $P_1$ and $P_2$ are beam polarizations, $L$ is the integrated
luminosity, $R$ is the relative luminosity and $N$ is the particle
yield.  In this report, we focus on $A_{LL}$ in $\pi^0$ production.

\section{Measurement}

A proton-carbon Coulomb Nuclear Interference (CNI) polarimeter [2]
was used on a fill-by-fill basis for measuring the relative
polarization.  For 2005 (2006), the average polarization was 47\%
(62\%).  In 2004, the absolute polarization in RHIC was measured
with a polarized hydrogen gas jet target \cite{RHIC:poljet},
reducing the relative uncertainty in the CNI value to 20\% per beam.
This leads to a 40\% scaling uncertainty for our final $A_{LL}$.
Finalized results from the jet target for 2005 and 2006 were not ready as of
this work.

In RHIC, the stable polarization direction is vertical, and so the
beam polarization must be rotated parallel to the beam momentum axis
when entering the PHENIX interaction area. By measuring the
amplitude of a single transverse spin asymmetry in forward neutrons
\cite{PHENIX:neutron}, the remaining transverse polarization
component can be determined.

\begin{figure}
  \includegraphics[height=.3\textheight]{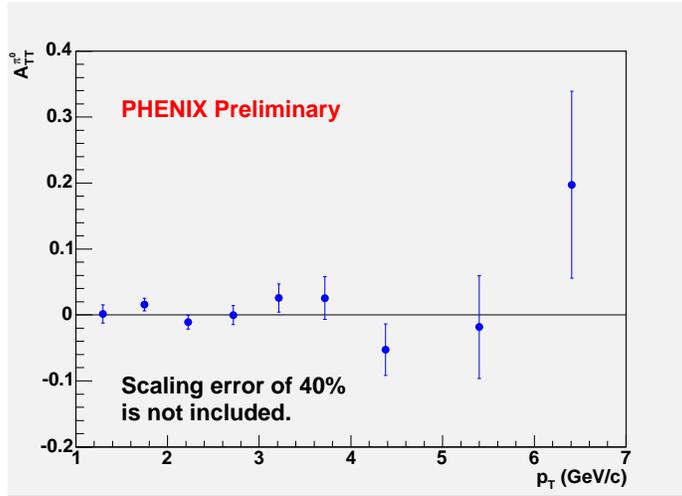}
  \caption{Azimuthally independent double transverse spin asymmetry,
  $A_{TT}$, in $\pi^0$ production during 2005 RHIC running period with
  transversely polarized protons.  Due to small remaining transverse beam
  polarizations in longitudinally polarized beams, $A_{TT}$ is a
  contamination in our measured asymmetry.  A systematic uncertainty of
  $0.075\times \delta A_{LL}$ is applied to the measured $A_{LL}^{\pi^0}$.}
  \label{p:ATT}
\end{figure}

The actual measured asymmetry is
\begin{equation}
A_{meas} = (1-\epsilon)A_{LL} + \epsilon A_{TT},\quad A_{TT} =
\frac{\sigma_{\uparrow\uparrow} -
\sigma_{\uparrow\downarrow}}{\sigma_{\uparrow\uparrow} +
\sigma_{\uparrow\downarrow}}\label{e:A_TT}
\end{equation}
where $A_{TT}$ is the azimuthally independent double transverse spin
asymmetry. (Note that $A_{LT}$ is parity violating.) $\epsilon$ is
the product of the remaining transverse spin components of the two
beams, which in 2005 was found to be 0.014. In the 2005 run, a small
data sample was taken with both beams transverse and was used in the
measurement of $A_{TT}$ is shown in Fig. \ref{p:ATT}. It was found
to be consistent with zero. Therefore, we assume a systematic
uncertainty in our $A_{LL}^{\pi^0}$ of 0.075$\times \delta
A_{LL}|_{stat}$.

To measure relative luminosity (Eq. \ref{A_LL_RL}), we use beam-beam
counters \cite{pi0:2003}. In PHENIX for the 2005 (2006) run,
uncertainty in relative luminosity was $\delta R=1.0\times10^{-4}$
($\delta R$~$=$~$1.1\times10^{-4}$) corresponding to a $\delta
A_{LL}|_{R} = 2.3\times10^{-4}$ ($\delta A_{LL}|_{R} =
1.5\times10^{-4}$).

The PHENIX Electromagnetic Calorimeter \cite{PHENIX:detector} is
specifically designed for very good energy and spacial resolution,
at the cost of limited acceptance ($|\eta|<0.35$, $\Delta
\phi=2\times 90^{\circ}$) and, in combination with a high $p_T$
photon trigger, is used to measure $\pi^0$ yield. The $\pi^0$
cross-section has been measured at PHENIX for a wide $p_T$ range
covering the $p_T$ range used in the $A_{LL}$ measurement. Comparing
NLO pQCD (using KKP fragmentation functions) to the data shows good
agreement for the entire $p_T$ range \cite{pi0:cross}. Therefore, we
can use NLO pQCD to interpret our measured $\pi^0$ $A_{LL}$ in terms
of $\Delta g$.

As of this work, the full data production for the 2006 run was not
completed.  However, events with a sufficiently high energy photon
were filtered during data taking for fast analysis.  Due to the
photon energy requirement, only data above $p_T=5$ GeV are shown.
Results from 2006 for $p_T<5$ GeV will be available when the whole
data set is produced.

\section{Asymmetry Calculation}

$A_{LL}$ is calculated using Eq. \ref{A_LL_RL} for the diphoton
invariant-mass range of $\pm25$ MeV/c$^2$ around the $\pi^0$ peak
($A_{LL}^{\pi^0+BG}$) and in two 50 MeV/c$^2$ wide mass regions on
either side ($A_{LL}^{BG}$).  Fitting the mass peak, we obtain
$w_{BG}$, the fraction of background in the peak region.  To obtain
$A_{LL}^{\pi^0}$, we use
\begin{equation}
A_{LL}^{\pi^0}=\frac{A_{LL}^{\pi^0+BG} -
w_{BG}A_{LL}^{BG}}{1-w_{BG}}.
 \label{e:ALLsubtraction}
\end{equation}

Bunch shuffling \cite{pi0:2003} was performed to evaluate systematic
uncertainty from bunch to bunch or fill to fill. For both the 2005
and 2006 runs, bunch-to-bunch and fill-to-fill systematic
uncertainty is negligible compared with our current statistical
uncertainty.

\section{$A_{LL}^{\pi^0}$ Results}

\begin{figure}
  \includegraphics[height=.3\textheight]{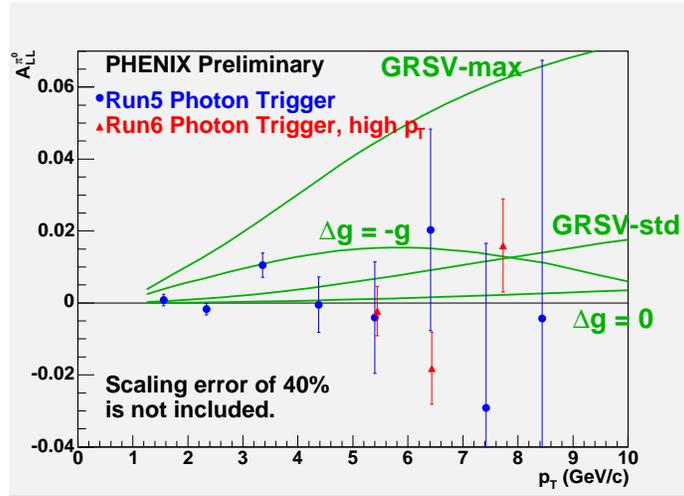}
  \caption{$\pi^0$ $A_{LL}$ results from 2005 RHIC run (blue circles) and 2006
  RHIC run (red triangles) plotted with four theory curves calculated with
  NLO pQCD \cite{pi0:GRSV}.  Curves are calculated with results from polarized DIS,
  assuming four values for gluon polarization described in the text.  Run6 result
  currently only includes high $p_T$ photon filtered data, with more data for
  $p_T<5$ GeV expected soon.}
  \label{p:ALL}
\end{figure}

Preliminary results from the 2005 RHIC run \cite{pi0:2005} are
plotted in Fig. \ref{p:ALL} (blue circles).  These data correspond
to an integrated luminosity ($L$) of 1.8 pb$^{-1}$ (~70\% of total
2005 luminosity) and average polarization ($\langle P \rangle$) of
47\%, giving a figure of merit ($\langle P\rangle ^4 L$) of 0.088
pb$^{-1}$. During the longitudinal polarization running period in
2006, PHENIX recorded $L=7.5$ pb$^{-1}$ with $\langle P
\rangle=62$\%, corresponding to a figure of merit 12.6 times larger
than that of the 2005 result.  Results from a fast track analysis on
a sub-sample of the 2006 data with a high $p_T$ photon are plotted
in Fig. \ref{p:ALL} (red triangles).

The data in Fig. \ref{p:ALL} are plotted with four theory curves
based on \cite{pi0:GRSV} assuming different input values of $\Delta
g$ at $Q^2$=0.4~GeV$^2$.  Three curves cover the full range of
possibilities: the maximal positive value $\Delta g$=$g$ (GRSV-max),
maximal negative value $\Delta g$=$-g$, and $\Delta g$=0, where $g$
is the unpolarized gluon parton distribution function. The fourth
curve, labeled GRSV-std, assumes a value of $\Delta g$ from the best
fit to the world DIS data as of \cite{pi0:GRSV}.

By calculating $\chi^2/NDF$ for the four theory curves using data
from 2005 and 2006, both positive and negative maximal scenarios are
found to have less than 0.1\% likelihood of fitting our data. Both
GRSV-std and $\Delta g=0$ are found to be consistent with our data,
with likelihood of 2-9\% and 11-12\%, respectively.  The range comes
from including the scaling uncertainty in beam polarization.
Theoretical uncertainties have not been taken into account.  To
remove possible soft physics influence, this test has been repeated
neglecting the data with $p_T<2$~GeV. Again, positive and negative
maximal gluon polarizations are strongly disfavored, while GRSV-std
(1-6\%) and $\Delta g=0$ (8-9\%) are not inconsistent with our data.

\section{Summary}

During the 2005 RHIC run, $A_{LL}^{\pi^0}$ was measured at
mid-rapidity and $\sqrt{s}$=200 GeV with the PHENIX detector.  2005
was the first significant run with polarized protons.  In 2006, the
figure of merit increased by a factor of 7.5.  Using a small subset
of the full 2006 data set with a high $p_T$ photon for a fast
analysis, $A_{LL}^{\pi^0}$ was presented for 5 GeV$<p_{T}<$9 GeV.
The 2005 and 2006 results strongly disfavor a large positive $\Delta
g$, and also disfavor a maximal negative $\Delta g$.  Both GRSV
standard and $\Delta g$=0 are consistent with our data.



\bibliographystyle{aipproc}   

\bibliography{spin2006}

\IfFileExists{\jobname.bbl}{}
 {\typeout{}
  \typeout{******************************************}
  \typeout{** Please run "bibtex \jobname" to optain}
  \typeout{** the bibliography and then re-run LaTeX}
  \typeout{** twice to fix the references!}
  \typeout{******************************************}
  \typeout{}
 }

\end{document}